\documentclass[10pt,journal]{IEEEtran} 
 \usepackage{booktabs} 
\usepackage{cite}
\usepackage{amsmath,amssymb,amsfonts}
\usepackage{algorithmic}
\usepackage{graphicx}
\usepackage{textcomp}
\usepackage{xcolor}
\usepackage{mathtools}
\def\BibTeX{{\rm B\kern-.05em{\sc i\kern-.025em b}\kern-.08em
    T\kern-.1667em\lower.7ex\hbox{E}\kern-.125emX}}

\usepackage{algorithm}
\usepackage{algorithmic}

\usepackage{color}
\usepackage{tablefootnote}

\usepackage{amsmath,amssymb,amsthm}
\usepackage{multirow}

\newcommand{\mbf}{\mathbf}

\usepackage{algorithm}
\usepackage{algorithmic}
\usepackage{longtable,tabularx}

\begin{document}

\title{Quadratic Extended and Unscented Kalman Filter Updates}

\author{

\IEEEauthorblockN{Simone Servadio,}
\IEEEauthorblockA{\textit{Iowa State University}, Ames, IA, USA}
\thanks{Dr. Simone Servadio, Assistant Professor, Department of Aerospace Engineering, servadio@iastate.edu}

\IEEEauthorblockN{Chiran. B. Cherian,}
\IEEEauthorblockA{\textit{Iowa State University}, Ames, IA, USA}
\thanks{Mr. Chiran Cherian, Graduate Student, Department of Aerospace Engineering, cbckbc@iastate.edu}

}


\maketitle
\thispagestyle{plain}
\pagestyle{plain}

\begin{abstract}
Common filters are usually based on the linear approximation of the optimal minimum mean square error estimator. The Extended and Unscented Kalman Filters handle nonlinearity through linearization and unscented transformation, respectively, but remain linear estimators, meaning that the state estimate is a linear function of the measurement. This paper proposes a quadratic approximation of the optimal estimator, creating the Quadratic Extended and Quadratic Unscented Kalman Filter. These retain the structure of their linear counterpart, but include information from the measurement square to obtain a more accurate estimate. Numerical results show the benefits in accuracy of the new technique, which can be generalized to upgrade other linear estimators to their quadratic versions. 
\end{abstract}

\IEEEpeerreviewmaketitle

\section{Introduction}
For linear, Gaussian systems, the posterior remains Gaussian, and the Kalman Filter \cite{kalman1960new} provides an exact, recursive mechanism to compute its mean and covariance. However, most systems exhibit nonlinear dynamics and/or measurements, leading to non-Gaussian probability density functions (PDFs).  A common way to handle such nonlinearities is to linearize the system about the current estimate. The Extended Kalman Filter (EKF) \cite{gelb1974applied} applies the Kalman mechanization to this linearized model. Despite its simplicity, the EKF often struggles to converge when nonlinearities are severe \cite{junkins2004nonlinear}.

An alternative is the Unscented Kalman Filter (UKF) \cite{julier2004unscented}, which uses a deterministic sampling to more accurately approximate the predicted mean and covariance. The unscented transformation approximates distributions as Gaussians and applies the nonlinear transformation of a set of well-selected sigma points. The UKF typically achieves higher accuracy and robustness than the EKF, since it accounts for higher-order terms in the nonlinear maps without explicitly computing Jacobians. Nevertheless, both EKF and UKF remain fundamentally linear estimators: at each step, the updated state is a linear function of the measurement. Both filters are derived from the linearized Minimum Mean Square Error (MMSE) principle, and they can be seen as different linear approximations of the conditional mean. Other ``linear” filters approximate the nonlinear dynamics and measurement equation in different ways: Cubature Kalman filters (CKF) \cite{arasaratnam2009cubature}, Ensemble Kalman filters (EnKF) \cite{servadio2021differential,valli2012gaussian}, Central Difference Kalman filters (CDKF) \cite{schei1997finite}. All of these still produce a linear measurement update, leaving the true MMSE estimator, which is generally nonlinear and intractable, unrealized. Even the use of State Transition Tensors (STT) \cite{majji2008high} and, originally, Differential Algebra \cite{valli2013nonlinear} have been used to improve the accuracy of the prediction step of the filters, on how they propagate and transform uncertainties, leaving the update structure untouched. 

To go beyond linear estimators, one can consider Gaussian Sum Filters (GSF) \cite{sorenson1971recursive, alspach2003nonlinear}, based on the concept of representing nongaussian distributions as a sum of multiple Gaussian components, or particle filters (PF), such as the bootstrap PF (BPF) \cite{gordon1993novel}, the Gaussian PF (GPF) \cite{hutter2003gaussian}, and other PF based on Sequential Importance Sampling (SIS) \cite{servadio2024likelihood}. However, these techniques leave the Kalman formulation to achieve an accurate estimate, based on the representation of PDFs via Gaussian Multiple Models (GMM) or as probability mass functions (PMF), yielding superior accuracy at the cost of significantly higher computational load.

A less explored route is to retain a linear update structure but replace the linear dependence on the measurement with a polynomial function. A nonlinear update within the Kalman structure has been offered in \cite{de1995optimal}, augmenting the measurement with its square to form a quadratic update, and this idea was extended to arbitrary polynomial orders \cite{carravetta1997polynomial}. This formulation requires carrying estimates of the state squares, and it presents flaws when working with vector measurements \cite{de1995optimal}. Indeed, the polynomial update presented in \cite{germani2005polynomial} has been tested only with scalar measurements, avoiding the issue of matrix singularity. Moreover, its derivation requires lifting the nonlinear system via a Carleman approximation, creating large and cumbersome matrices.

In more recent studies, a different quadratic update (extendable to a polynomial update of any order) based on Taylor series expansions has been derived \cite{servadio2020recursive}. It requires knowledge of high-order central moments, calculated in the Differential Algebra (DA) framework, and carried at each time step, likewise the EKF carries mean and covariance. Further development \cite{servadio2020nonlinear} performs a polynomial update without carrying the higher-order central moments and, thus, reduces the overall computational cost. In \cite{servadio2020nonlinear}, non-Gaussian distributions are approximated as a polynomial transformation of Gaussian random variables, where moments are evaluated efficiently in DA using Isserlis' formulation \cite{isserlis1918formula}. Lastly, the polynomial update has been expanded to the estimation of the conditional covariance, applying a double update sequentially for mean and covariance \cite{servadio2021estimation}.

This paper offers the derivation of two estimators based on the quadratic approximation of the MMSE, achieved via linearization and through the unscented transformation. The proposed Quadratic Extended Kalman Filter (QEKF) and Quadratic Unscented Kalman Filter (QUKF) closely resemble the EKF and the UKF structure, but they apply augmented covariances to achieve a nonlinear measurement update, where the state estimate is a parabolic function of the measurement. Numerical results show how the newly proposed nonlinear estimators outperform their linear counterparts.

\section{The Optimal Quadratic Estimator} \label{sec1}
Previous works \cite{servadio2020recursive, servadio2020nonlinear, servadio2021estimation} derived a new series of polynomial estimators where the state estimate is a polynomial function of the given measurement. They make use of Differential Algebra techniques to calculate high-order central moments. In this work, we obtain the quadratic estimator from the definition of the orthogonality principle, but apply linearization and the unscented transformation to obtain the high-order gains and terms, so that the final results resemble the structure of the Extended and Unscented Kalman Filters, but with the quadratic term. 

Minimum Mean Square Error (MMSE) estimators aim at minimizing the expected value of the error square; thus, their formulation leads to unbiased filters where the mean of the error square coincides with the error covariance. Whenever a linear function in the measurement is used, as a family of estimators, we obtain a linear representation of the MMSE, the LMMSE. These estimators have the form of
\begin{equation}
  g_L(\mbf y) = A + \mathbb{E}[\mbf x] + B\,(\mbf y - \mathbb{E}[\mbf y])  \label{eq:linest}
\end{equation}
and are very common. Equation \eqref{eq:linest} is used by the EKF, the UKF, and other linear estimators in their state update step. Their estimate changes according to the technique used to evaluate the constants $A$ and $B$, dealing with the nonlinearities of the measurement model. 

This work applies a quadratic approximation to the MMSE, the QMMSE. Thus, the generic quadratic estimator family is
\begin{align}
  g_Q(\mbf y) &= A + \mathbb{E}[\mbf x] +B\delta \mbf y + C\delta \mbf y ^{[2]}
\end{align}
where
\begin{align}
  \delta \mbf y &= \mbf y - \mathbb{E}[\mbf y] \\
  \delta \mbf y ^{[2]} &= \delta \mbf y \otimes\delta \mbf y = (\mbf y - \mathbb{E}[\mbf y]) \otimes (\mbf y - \mathbb{E}[\mbf y]) 
\end{align}
indicates the residual (or measurement deviation vector) and its square, obtained via the Kronecker product $\otimes$ for the square of vectors. 

The orthogonality principle states that, given the optimal estimator $g^*_Q(\mbf y)$, with optimal constants $A^*$, $B^*$, and $C^*$, the equation
\begin{equation}
  \mathbb{E}\bigl[(\mbf x - g^*_Q(\mbf y))\,g_Q(\mbf y)^T\bigr] = 0 \label{eq:ortho}
\end{equation}
holds for every other generic estimator. Therefore, picking 
\begin{equation}
A =- \mathbb{E}[\mbf x] + \mbf I, \quad B = \mbf 0, \quad C = \mbf 0 
\end{equation}
where $\mbf I$ is the identity matrix, leads to
\begin{align}
  \mathbb{E}[\delta \mbf x - A^* - B^*\delta \mbf y - C^*\delta \mbf y^{[2]}] =&  \nonumber\\
   A^* =& -C^*\,\mathrm{v}(\mbf P_{yy}). \label{eq:3}
\end{align}
where $\mathrm{v}(\mbf P_{yy})$ indicates the stack operator for matrix $\mbf P_{yy}$.
In a similar way, selecting the two sets 
\begin{align}
A &=- \mathbb{E}[\mbf x], \quad B = \mbf I, \quad C = \mbf 0  \\
A &=- \mathbb{E}[\mbf x], \quad B = \mbf 0, \quad C = \mbf I
\end{align}
leads to 
\begin{align}
\mathbb{E}[(\delta \mbf x - A^* - B^*\delta \mbf y - C^*\delta \mbf y^{[2]})\,\delta \mbf y^T]=& \nonumber \\
  \mbf P_{xy} - B^*\mbf P_{yy} - C^*\mbf P_{y^{[2]}y} =& 0\label{eq:1}
\end{align}
and
\begin{align}
\mathbb{E}[(\delta \mbf x - A^* - B^*\delta \mbf y - C^*\delta \mbf y^{[2]})\,\delta \mbf y^{[2]T}]&= \nonumber\\ 
  \mbf P_{xy^{[2]}} - B^*\mbf P_{yy^{[2]}} + C^*(\mathrm{v}(\mbf P_{yy})\mathrm{v}(\mbf P_{yy})^T-\mbf P_{y^{[2]}y^{[2]}}) &= 0 \label{eq:2}
\end{align}
where we have defined covariances as 
\begin{align} \label{eq:set_in}
  \mbf P_{xy} &= \mathbb{E}[\delta \mbf x\,\delta \mbf y^T],\\
  \mbf P_{xy^{[2]}} &= \mathbb{E}[\delta \mbf x\,\delta \mbf y^{[2]^T}],\\
  \mbf P_{yy} &= \mathbb{E}[\delta \mbf y\,\delta \mbf y^T],\\
  \mbf P_{yy^{[2]}} &= \mathbb{E}[\delta \mbf y\,\delta\mbf  y^{[2]^T}],\\
  \mbf P_{y^{[2]}y} &= \mathbb{E}[\delta \mbf y^{[2]}\,\delta \mbf y^T],\\
  \mbf P_{y^{[2]}y^{[2]}} &= \mathbb{E}[\delta \mbf y^{[2]}\,\delta \mbf y^{[2]^T}].\label{eq:set_fin}
\end{align}
with
\begin{align}
  \delta \mbf x &= \mbf x - \mathbb{E}[\mbf x] 
\end{align}
After substituting in Eq. \eqref{eq:3}, Eqs \eqref{eq:1} and \eqref{eq:2} can be rewritten as a set of coupled equations.
\begin{equation}
  \begin{bmatrix} B^* & C^* \end{bmatrix}
  \begin{bmatrix}
    \mbf P_{yy} & \mbf P_{yy^{[2]}} \\
    \mbf P_{y^{[2]}y} & \mbf P_{y^{[2]}y^{[2]}} - \mathrm{v}(\mbf P_{yy})\mathrm{v}(\mbf P_{yy})^T
  \end{bmatrix}
  =
  \begin{bmatrix} \mbf P_{xy} \\ \mbf P_{xy^{[2]}} \end{bmatrix}^T
\end{equation}
which brings to the solution for the optimal constants
\begin{equation}
  \mbf K = \begin{bmatrix} B^* & C^* \end{bmatrix} =  \mbf P_{x \mathcal Y}\mbf P_{\mathcal Y \mathcal Y}^{-1}\label{eq:4}
\end{equation}
having defined 
\begin{align}
  \mbf P_{x \mathcal Y} &=  \begin{bmatrix} \mbf P_{xy} & \mbf P_{xy^{[2]}} \end{bmatrix} \\
  \mbf P_{\mathcal Y \mathcal Y} &= \begin{bmatrix}
  \mbf P_{yy} & \mbf P_{yy^{[2]}}\\ \mbf P_{y^{[2]}y} & \mbf P_{y^{[2]}y^{[2]}} - \mathrm{v}(\mbf P_{yy})\mathrm{v}(\mbf P_{yy})^T
  \end{bmatrix}
\end{align}
The optimal constants $B^*$ and $C^*$ behave likewise the classic Kalman gain, but evaluated in its augmented form that considers the covariances and cross covariances with the square of the measurement deviation vector. 

Substituting the derived constants from Eq. \eqref{eq:3} and \eqref{eq:4} back into the definition of the quadratic estimator, it creates the optimal quadratic estimator, which is the parabolic approximation of the true MMSE.
\begin{equation}
  g^*_Q(\mbf y) = \mathbb{E}[\mbf x] + \mbf P_{x \mathcal Y}\mbf P_{\mathcal Y \mathcal Y}^{-1} 
  \begin{bmatrix}\delta \mbf y \\
          \delta \mbf y^{[2]} - \mathrm{v}(\mbf P_{yy})\end{bmatrix}.
\end{equation}

The derived estimator is applied in the update step of the classic Extended Kalman Filter (EKF) and Unscented Kalman Filter (UKF) to obtain the quadratic version of their formulation, which is linear. Therefore, using linearization to evaluate the set of central moments in Eqs \eqref{eq:set_in} to \eqref{eq:set_fin}, we propose the Quadratic Extended Kalman Filter (QEKF) and, using the unscented transformation to evaluate moments, the Quadratic Unscented Kalman Filter (UKF).

\section{High Order Central Moments}
Linear estimators require the first two central moments, i.e., mean and covariance, to operate. On the contrary, quadratic estimators, due to their higher order, require information up to the fourth central moment to achieve a correct and consistent estimate. Therefore, we define with 
\begin{equation}
    \mbf S_{xxx} = \mathbb{E}\bigl[(x - \hat{x}) \otimes \big((x - \hat{x})(x - \hat{x})^T\big)\bigr]
\end{equation}
the skewness of a distribution, expressed as a three-dimensional tensor, and with 
\begin{equation}
    \mbf K_{xxxx} = \mathbb{E}\bigl[\big((x - \hat{x})(x - \hat{x})\big)^T \otimes \big((x - \hat{x})(x - \hat{x})^T\big)\bigr]
\end{equation}
the kurtosis, expressed as a forth dimensional tensor. It can be shown \cite{servadio2020nonlinear, servadio2021estimation} that, for a null mean distribution, where central moments and raw moments match, $\mbf S_{xxx} = \mbf P_{x^{[2]}x} = \mbf P_{xx^{[2]}}^T $ and $\mbf K_{xxxx} = \mbf P_{x^{[2]}x^{[2]}}$.

\section{The Quadratic Extended Kalman Filter (QEKF)}
The formulation of the EKF is well-known: it linearizes the dynamic and measurement model at the current estimate and it applies the Kalman Filter algorithm \textit{as-if} everything is linear and Gaussian. It represents distributions as Gaussians, propagating and updating, iteratively, the state mean, $\hat {\mbf x}$, and covariance, $\mbf P_{xx}$.

Consider the nonlinear equation of motion, with dynamics $f()$, affected by noise
\begin{equation}
    \mbf x_{k} = f(\mbf x_{k-1}) + \boldsymbol{\mu}_{k-1}
\end{equation}
where $\mu_{k-1}$ is a zero-mean Gaussian process noise with known covariance matrix $\mbf P_{\mu\mu}$. According to the EKF implementation, the prediction step propagates mean and covariance according to 
\begin{align}
    \hat{\mbf x}^-_{k} &= f(\hat{\mbf x}^+_{k-1}) \quad \\
    \mbf P_{xx,k}^- &= \mbf F_{k-1} \mbf P_{xx,k-1}^+ \mbf F_{k-1}^T+\mbf P_{\mu\mu}\\
    \mbf F_{k-1} &= \left. \frac{\partial f}{\partial \mbf x} \right|_{\hat{\mbf x}^+_{k-1}}
\end{align}
where $ \hat{\mbf x}^+_{k-1}$ is the state mean at the previous time step, $ \hat{\mbf x}^-_{k}$ is the current predicted mean of the distribution, $\mbf P_{xx}^-$ indicates the predicted (prior) covariance, and $\mbf F_k$ is the gradient of the dynamics. 

The state PDF is now approximated as a Gaussian with know mean and covariance, which constitutes the prior for the Bayesian update. An observation is provided according to the measurement model $h()$,
\begin{equation}
  \mbf y = h(\mbf x) + \boldsymbol{\eta}
\end{equation}
affected by zero-mean measurement noise with known covariance $\mbf P_{\eta\eta}$, skewness $\mbf S_{\eta\eta\eta}$, and kurtosis $\mbf K_{\eta\eta\eta\eta}$. The predicted measurement mean is 
\begin{equation}
  \hat{\mbf y}_k = h(\hat{\mbf x}^-_{k}) 
\end{equation}
Given the measurement outcome $\tilde {\mbf y}$, the actual numerical value of the random value from the sensors, the residual and its square are evaluated as 
\begin{align}
    \delta \tilde{\mbf y} &= \mbf y - \hat{\mbf y}_k \\
  \delta\tilde{\mbf y}^{[2]} &= \delta \tilde{\mbf y}\otimes\delta \tilde{\mbf y} 
\end{align}
so that the update step of the QEKF can be performed according to 
\begin{align}
        \mbf K &= \mbf P_{x \mathcal Y}\mbf P_{\mathcal Y \mathcal Y}^{-1} \\
     \hat{\mbf x}^+_{k}&= \hat{\mbf x}^-_{k} +\mbf K
  \begin{bmatrix}\delta \mbf y \\
          \delta \mbf y^{[2]} - \mathrm{v}(\mbf P_{yy})\end{bmatrix} \label{eq:up1}\\
          \mbf P_{xx,k}^+&=\mbf P_{xx,k}^--\mbf K\mbf P_{\mathcal Y \mathcal Y}\mbf K^T \label{eq:up2}
\end{align}
where $\mbf K $ is the Kalman gain, in its augmented form, and $\hat{\mbf x}^+_{k}$ and $\mbf P_{xx,k}^+$ are the updated mean and covariance of the state distribution. 

The reader can notice how Eqs \eqref{eq:up1} and \eqref{eq:up2} resemble the structure of the normal Kalman update, but expanded to operate with the augmented covariances that include information of the square of the measurement vector. Indeed, the state-measurement cross-covariance is evaluated block-wise as in 
\begin{equation}
    \mbf P_{x \mathcal Y} =  \begin{bmatrix} \mbf P_{xy} & \mbf P_{xy^{[2]}} \end{bmatrix}  \label{eq:pxy}
\end{equation}
with 
\begin{align}
    \mbf P_{x y} &= \mbf P_{xx,k}^- \mbf H^T\\
    \mbf P_{xy^{[2]}} &= \mbf S_{xxx}\mbf H^{[2]T}\\
    \mbf H_k &= \left. \frac{\partial h}{\partial \mbf x} \right|_{\hat{\mbf x}^-_{k}}
\end{align}
where $\mbf H^{[2]} = \mbf H \otimes \mbf H$ squares the measurement model. In a similar pattern, the augmented measurement covariance is evaluated block-wise as in 
\begin{equation}
    \mbf P_{\mathcal Y \mathcal Y} = \begin{bmatrix}
    \mbf P_{yy} & \mbf P_{yy^{[2]}} \\
    \mbf P_{yy{[2]}}^T & \mbf P_{y^{[2]}y^{[2]}} - \mathrm{v}(\mbf P_{yy})\mathrm{v}(\mbf P_{yy})^T
  \end{bmatrix}\label{eq:pyy}
\end{equation}
with 
\begin{align}
    \overline{\mbf P_{y y}} &= \mbf H \mbf P_{xx,k}^- \mbf H^T \nonumber \\
    \mbf P_{y y} &= \overline{\mbf P_{y y}} + \mbf P_{\eta\eta}\\
    \mbf P_{yy^{[2]}} &= \mbf H \mbf S_{xxx}\mbf H^{[2]T} + \mbf S_{\eta\eta\eta}\\
    \mbf P_{yy^{[2]}}  &=  \mbf H^{[2]} \mbf K_{xxxx} \mbf H^{[2]^T} + \mbf K_{\eta\eta\eta\eta} + \nonumber \\
    + &\overline{\mbf P_{y y}} \otimes \mbf P_{\eta\eta} + \mathrm{m}\Big( \mathrm{v}(\overline{\mbf P_{y y}}) \otimes \mbf I  \Big)\Big( \mbf P_{\eta\eta}^T \otimes \mbf I \Big) + \nonumber \\
    + & \mbf P_{\eta\eta} \otimes \overline{\mbf P_{y y}}+  
    \mathrm{m}\Big( \mathrm{v}(\mbf P_{\eta\eta}) \otimes \mbf I  \Big)\Big( \overline{\mbf P_{y y}}^T \otimes \mbf I \Big) +\nonumber\\
    + &\mbf H^{[2]}\mathrm{v}(\mbf P_{xx})\mathrm{v}(\mbf P_{\eta\eta})^T + \mathrm{v}(\mbf P_{\eta\eta})\mathrm{v}(\mbf P_{xx})^T\mbf H^{[2]T} 
\end{align}
where $\mathrm{m}()$ is the matrix operator, inverse of the stack operator $\mathrm{v}()$, and where $\overline{\mbf P_{y y}}$ indicates the measurement covariance without any influence from the noise, the actual uncertainty transformation.  

These expressions of the covariances are obtained by applying the expected value operator as described in Eqs \eqref{eq:set_in} to \eqref{eq:set_fin} with the linearized model of the measurement equation. Full knowledge of the moments of the noise is assumed, while the state prior skewness and kurtosis can be approximated in a way that better fits the application. For a simple Gaussian assumption, each entry of the tensors can follow the Isserlis' formulation \cite{isserlis1918formula}:
\begin{align}
    \mbf S_{ijk} &= 0 \\
    \mbf K_{ijkm} &= \mbf P_{ij}\mbf P_{km} + \mbf P_{ik}\mbf P_{jm} + \mbf P_{im}\mbf P_{jk} 
\end{align}

The state has been updated according to  Eqs \eqref{eq:up1} and \eqref{eq:up2} and the mean and covariance can undergo a new propagation normally. If propagation is of importance, a higher accuracy prediction step such as DA or STT can be applied to obtain the prior state distribution, followed by the quadratic update proposed here.

\section{The Quadratic Unscented Kalman Filter (QUKF)}
The quadratic estimator derived in Section II is generic and can be applied to multiple linear filters, to expand the technique that deals with the system and equation nonlinearity in the prediction to achieve a nonlinear update. In this Section, we exploit the unscented transformation to obtain the QUKF. 
The unscented transformation (UT) is applied to obtain the mean and covariance of a random variable, $\mbf y$, which has a known relation, $h()$, with respect to the given initial random variable $\mbf x$, of dimension $n$. After defining the spread parameter $\alpha$, the scaling parameter $\kappa$, and the constant $\beta$, a set of  $2n+ 1$ sigma points is created from the current state mean and covariance: 
\begin{align}
    \lambda &= \alpha^2 (n + \kappa) - n \\
    \mbf C_{k-1}\mbf C_{k-1}^T &=  (n+\lambda)\mbf P_{xx,k-1}^+\\
    \boldsymbol{\chi}_{k-1} &= \begin{bmatrix} \hat{\mbf x}^+_{k-1}  & \hat{\mbf x}^+_{k-1} +\mbf C_{k-1}   &\hat{\mbf x}^+_{k} -\mbf C_{k-1} \end{bmatrix}
\end{align}
Each sigma point is associated with a weight, given by
\begin{align}
    w_0^{(m)} &= \frac{\lambda}{n + \lambda} \\
    w_0^{(c)} &= \frac{\lambda}{n + \lambda} + (1 - \alpha^2 + \beta) \\
    w_i^{(m)} &= w_i^{(c)} = \frac{1}{2(n + \lambda)}, \quad i = 1, \dots, 2n
\end{align}
for the calculation of means, $w_i^{(m)}$, and covariances, $w_i^{(c)}$. 

The prediction step is performed by propagating, separately, each sigma point and obtaining their weighted means for the predicted estimate and covariance. 
\begin{align}
    \boldsymbol{\chi}_{k}^{(i)} &= f(\boldsymbol{\chi}_{k-1}^{(i)}) \quad \forall i = 0,...,2n \\
    \hat{\mbf x}^-_{k} &=  \sum^{2n}_{i=0} w^{(i)}_{m}\boldsymbol{\chi}_{k}^{(i)}  \\
    \mbf P_{xx,k}^- &= \sum^{2n}_{i=0} w^{(i)}_{c} (\boldsymbol{\chi}_{k}^{(i)}-\hat{\mbf x}^-_{k})(\boldsymbol{\chi}_{k}^{(i)}-\hat{\mbf x}^-_{k})^T     +\mbf P_{\mu\mu}
\end{align}

The quadratic update for the QUKF follows the same formulation as in Eqs \eqref{eq:up1} and \eqref{eq:up2}, with the augmented covariances reported in Eqs \eqref{eq:pxy} and \eqref{eq:pyy}, evaluated block-wise. However, each entry is computed via the sigma points rather than with linearization. Therefore, after obtaining the transformed sigma points in the measurement space and their predicted mean
\begin{align}
    \boldsymbol{\mathcal Y}_{k}^{(i)} &= h(\boldsymbol{\chi}_{k}^{(i)}) \quad \forall i = 0,...,2n \\
    \hat{\mbf y}_{k} &= \sum^{2n}_{i=0} w^{(i)}_{m}\boldsymbol{\mathcal Y}_{k}^{(i)}
\end{align}
each sigma point is augmented with its Kronecker square. Thus, the deviation vectors of the state and the measurement are computed as 
\begin{align}
    \delta \boldsymbol{\chi}_{k}^{(i)} &= \boldsymbol{\chi}_{k}^{(i)} - \hat{\mbf x}^-_{k} \quad \forall i = 0,...,2n \\
    \delta \boldsymbol{\mathcal Y}_{k}^{(i)} &=\boldsymbol{\mathcal Y}_{k}^{(i)} -  \hat{\mbf y}_{k} \quad \forall i = 0,...,2n \\
    \delta \boldsymbol{\mathcal Y}_{k}^{(i)} &= \delta \boldsymbol{\mathcal Y}_{k}^{(i)} \otimes \delta \boldsymbol{\mathcal Y}_{k}^{[2](i)} \quad \forall i = 0,...,2n
\end{align}
This definition eases the evaluation of the covariances, which are obtained as simple summation. The state-measurement cross covariances components are
\begin{align}
    \mbf P_{x y} &= \sum^{2n}_{i=0} w^{(i)}_{c}\delta \boldsymbol{\chi}_{k}^{(i)}\delta \boldsymbol{\mathcal Y}_{k}^{(i)T} \label{eq:cv1}\\
    \mbf P_{xy^{[2]}} &= \sum^{2n}_{i=0} w^{(i)}_{c} \delta \boldsymbol{\chi}_{k}^{(i)} \delta \boldsymbol{\mathcal Y}_{k}^{[2](i)T}
\end{align}

The measurement covariance is evaluated in two separate steps: first, the transformed covariances in the measurement state are computed, then the influence of the noise in added. 
\begin{align}
    \overline{\mbf P_{y y}} &= \sum^{2n}_{i=0} w^{(i)}_{c}\delta \boldsymbol{\mathcal Y}_{k}^{(i)} \delta \boldsymbol{\mathcal Y}_{k}^{(i)T}\\
    \overline{\mbf P_{yy^{[2]}}} &= \sum^{2n}_{i=0} w^{(i)}_{c} \delta\boldsymbol{\mathcal Y}_{k}^{(i)} \delta \boldsymbol{\mathcal Y}_{k}^{[2](i)T} \\
    \overline{\mbf P_{y^{[2]}y^{[2]}}} &=\sum^{2n}_{i=0} w^{(i)}_{c} \delta \boldsymbol{\mathcal Y}_{k}^{[2](i)} \delta \boldsymbol{\mathcal Y}_{k}^{[2](i)T}
\end{align}
where additive noise leads to a simple noise addition for the contribution:
\begin{align}
    \mbf P_{y y} &= \overline{\mbf P_{y y}} + \mbf P_{\eta\eta}\\
    \mbf P_{yy^{[2]}} &= \overline{\mbf P_{yy^{[2]}}} + \mbf S_{\eta\eta\eta}\\
    \mbf P_{y^{[2]}y^{[2]}}  &=  \overline{\mbf P_{y^{[2]}y^{[2]}}} + \mbf K_{\eta\eta\eta\eta} + \nonumber \\
    + &\overline{\mbf P_{y y}} \otimes \mbf P_{\eta\eta} + \mathrm{m}\Big( \mathrm{v}(\overline{\mbf P_{y y}}) \otimes \mbf I  \Big)\Big( \mbf P_{\eta\eta}^T \otimes \mbf I \Big) + \nonumber \\
    + & \mbf P_{\eta\eta} \otimes \overline{\mbf P_{y y}}+  
    \mathrm{m}\Big( \mathrm{v}(\mbf P_{\eta\eta}) \otimes \mbf I  \Big)\Big( \overline{\mbf P_{y y}}^T \otimes \mbf I \Big) +\nonumber\\
    + &\mathrm{v}(\overline{\mbf P_{y y}})\mathrm{v}(\mbf P_{\eta\eta})^T + \mathrm{v}(\mbf P_{\eta\eta})\mathrm{v}(\overline{\mbf P_{y y}})^T\label{eq:cv2}
\end{align}

The QUKF algorithm is complete, as it has the same structure as the QEKF, where the difference lies in how the central moments (mean and covariances) are evaluated. The higher-order moments provide additional information to the filter with respect to its linear version, drastically improving the accuracy of the measurement update. 

\section{Generalization to Quadratic Updates}
A quick consideration about similarities and differences between the QEKF and the QUKF is due. The two filters have the same structure, similarly to how both the EKF and UKF follow the Kalman formulation. They create an augmented system that follows the same procedure regardless of the uncertainty transformation technique selected: linearization for the QEKF and unscented transformation for the QUKF.

Therefore, the proposed quadratic update fits any technique, and it can be applied to other linear estimators, such as the Cubature Kalman Filter or the Ensemble Kalman Filter, where it changes how Eqs \eqref{eq:cv1} to \eqref{eq:cv2} are evaluated, while keeping the same structure for the augmented squaring of the measurement update.

\section{Numerical Applications}
The proposed algorithms have been applied to two different applications: first, the newly developed update techniques have been used in a simple scalar toy problem to visually highlight the benefits of the quadratic approach; secondly, a more complex relative navigation in space application with non-Gaussian measurement noises has been tested.

\subsection{Scalar Problem}
A scalar problem is formulated to verify the improvements of the quadratic update, enabling nonlinear estimation, over any linear filtering techniques, such as the EKF and the UKF. The problem will show how the quadratic representation (QMMSE) better approximates the true MMSE function, especially when compared to its linear counterpart, the LMMSE.

Define a prior state $x \sim \mathcal{N}(1,0.05)$ and a measurement
\begin{align}
y = \mathrm{arctan}(x) + \eta
\end{align}
where $\eta \sim \mathcal{N}(0,0.01)$ is the measurement noise.

\begin{figure}  [htbp!]
    \centering
    \includegraphics[width=1\linewidth]{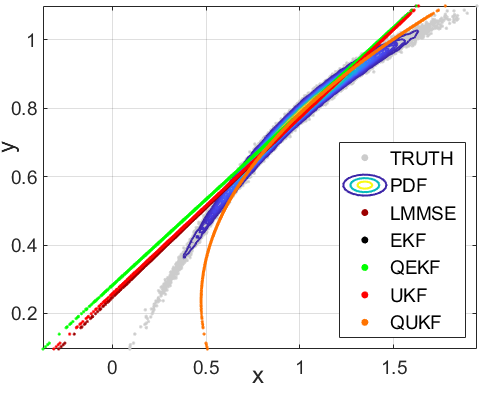}
    \caption{ Representation of the state-measurement joint distribution and of the relative estimates from different estimators.}
    \label{fig:post_est}
\end{figure}

We then simulate the true joint distribution of $x$ and $y$ using $10^5$ samples, reported in Fig. \ref{fig:post_est}. The optimal (nonlinear) MMSE is the conditional mean, which visually is the curved line dividing in half the distribution of $y$ (horizontal spread of grey points) for each value of $x$. Starting from a Gaussian distribution, the spread of the points follows the joint PDF (thus, the posterior) shape due to the nonlinearity of the measurement equation. The figure also shows different estimators in various colors: the EKF, UKF, QEKF, and QUKF. The EKF (black) and the UKF (red) are straight lines, whose slope is the Kalman gain. A more accurate transformation of central moments leads to a better prediction in the measurement space and, therefore, to a more reliable Kalman gain. Conceptually, the optimal Kalman gain obtainable is represented by the brown line, which represents the true LMMSE evaluated directly from the particles. Linear estimators, regardless of the technique that can be used to approximate moments, asymptotically aim at reaching this line. 
On the contrary, the quadratic estimators are given the capability of obtaining a curved function. In the figure, the QEKF overlaps the EKF. This is due to the Gaussian prior assumption, which leads to $P_{xy^2} = 0$, meaning the square term in the estimator function, $ay^2$, has $a = 0$, and it reduces to the EKF. This happens because a full Gaussian assumption is applied, supported by the Isserlis' formula. On the contrary, the QUKF calculates $P_{xy^2}$ directly from the sigma points, giving a non-null skewed behavior. The result is an estimator line that follows a parabolic trend in the variable $y$, as shown by the orange line. The QUKF estimate turns, and it follows the curved shape of the posterior distribution, achieving a more accurate estimate. The QUKF curved behavior improves both the representation of the tails of the distribution and the region around the mode of the posterior PDF. 

\begin{figure}  [htbp!]
    \centering
    \includegraphics[width=1\linewidth]{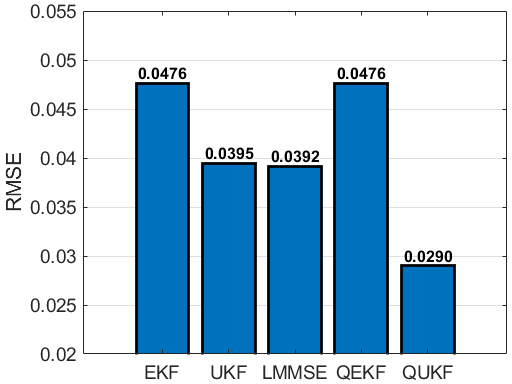}
    \caption{RMSE among selected estimators}
    \label{fig:rmse}
\end{figure}

The advantages of the quadratic approximation of the MMSE over the linear counterpart can be appreciated in Fig. \ref{fig:rmse}, where the root mean square error has been evaluated as
\begin{equation}
    RMSE = \sqrt{ \sum_{i=1}^N (\hat{x_i} - x_T)^2
    }
\end{equation}
where $x_T$ is the true value of $x$ and $N$ is the number of points representing the distribution. As previously mentioned, due to the Gaussianity assumption of the skewness, the QEKF reduces to the EKF, and they share the same error level. The UKF is more accurate than the EKF, improving accuracy drastically thanks to the unscented transformation being more reliable than mere linearization for the transformation of central moments. However, the UKF cannot surpass the limit given by the optimal LMMSE, which represents the most accurate linear estimator achievable. The LMMSE bar represents the best possible linear fit of the particles that compose the posterior distribution. 
On the other hand, the QUKF is able to better follow the shape of the distribution, and its error is much lower than the UKF, with an improvement of around 25\% in accuracy. Ref \cite{servadio2020nonlinear} showed that, conceptually, by increasing the order of the polynomial update to cubic and higher, the accuracy keeps improving, until reaching, asymptotically, the true MMSE. The QUKF derived in this paper enhance to use of the unscented transformation to obtain a quadratic polynomial representation of the true MMSE.

\subsection{Clohessy--Wiltshire Relative Motion}
The Clohessy--Wiltshire (CW) equations, also known as the Hill's equations, describe the relative motion of a chaser spacecraft with respect to a target spacecraft in a circular orbit. Assuming the target is in a circular orbit and a local-vertical-local-horizontal (LVLH) coordinate frame centered on the target, the linearized equations of motion for the relative position vector $\mathbf{r} = [x, y, z]^T$ are given by:
\begin{align}
    \ddot{x} &= 2n\dot{y} + 3n^2x \\
    \ddot{y} &= -2n\dot{x} \\
    \ddot{z} &= -n^2z
\end{align}
where $n$ is the mean motion of the chief's circular orbit, defined as \( n = \sqrt{\mu / a^3} \), with $\mu$ the Earth's standard gravitational parameter, and $a$ the semi-major axis of the chief's orbit, assumed to be 7000 km. These equations assume linearized dynamics, valid for small separations relative to the orbital radius. The initial relative PDF is assumed to be a Gaussian distribution with mean $\mathbf{x}_0 = \begin{bmatrix}
\mathbf{r}_0 &
\mathbf{v}_0 
\end{bmatrix} ^T$ 
given by
\begin{align}
\mathbf{r}_0 &= 
\begin{bmatrix}
2 &
10 &
-3.5 
\end{bmatrix} ^T \quad \text{km}\\
\mathbf{v}_0 &= 
\begin{bmatrix}
0.01 &
-0.005 &
0.0005
\end{bmatrix} ^T \quad \text{km/s}
\end{align}
and covariance \( \mathbf{P}_0 \) defined as:
\begin{equation}
\mathbf{P}_0 = \mathrm{blkdiag}\left(10^{-4}\mathbf{I}_3\text{km}^2,10^{-9}\mathbf{I}_3\text{km/s}^2\right)
\end{equation} 
where $\mathrm{blkdiag}()$ indicates the block diagonal operator and $\mathbf{I}_3$ a 3x3 identity matrix. 

Relative angle measurements are acquired by the chief every minute for 3 hours according to 
\begin{align}
y_1 &= \mathrm{arctan}(y/x) + \eta_1 \\
y_2 &= \mathrm{arcsin}\left(\dfrac{z}{\sqrt{x^2+y^2+z^2}}\right) + \eta_2
\end{align}
where $\eta_i$ is a zero-mean non-Gaussian noise, whose distribution is described in Table \ref{tab:1} \cite{servadio2020recursive,carravetta1997polynomial}. 

\begin{table}[h!]
\centering
\caption{Measurement noise distribution (rad)}
\begin{tabular}{c| c c c}
\toprule
$\eta_k$ & $1\mathrm{e}{-3}$ & $-3\mathrm{e}{-3}$ & $-9\mathrm{e}{-3}$ \\
\midrule
$P(\eta_k)$ & $\dfrac{15}{18}$ & $\dfrac{2}{18}$ & $\dfrac{1}{18}$ \\
\bottomrule
\end{tabular}
\label{tab:1}
\end{table}

\begin{figure}  [htbp!]
    \centering
    \includegraphics[width=.80\linewidth]{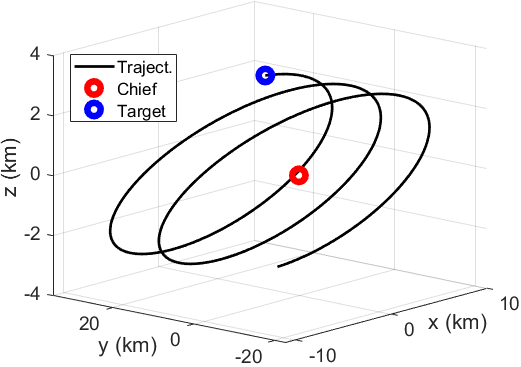}
    \caption{Target's relative motion with respect to the chief satellite.}
    \label{fig:traj}
\end{figure}

This application has been chosen since it presents a linear dynamic with nonlinear measurements. The linear system guarantees the correct uncertainty propagation of the state distribution, since a Gaussian PDF remains Gaussian after a linear transformation. Therefore, the difference and improvement in accuracy are due to the different order of the measurement update. Figure \ref{fig:traj} shows the target's relative motion with respect to the chief, creating a helical pattern around the origin whose amplitude is connected to the mean motion of the reference orbit. 

\begin{figure*}  [htbp!]
    \centering
    \includegraphics[width=0.93\linewidth]{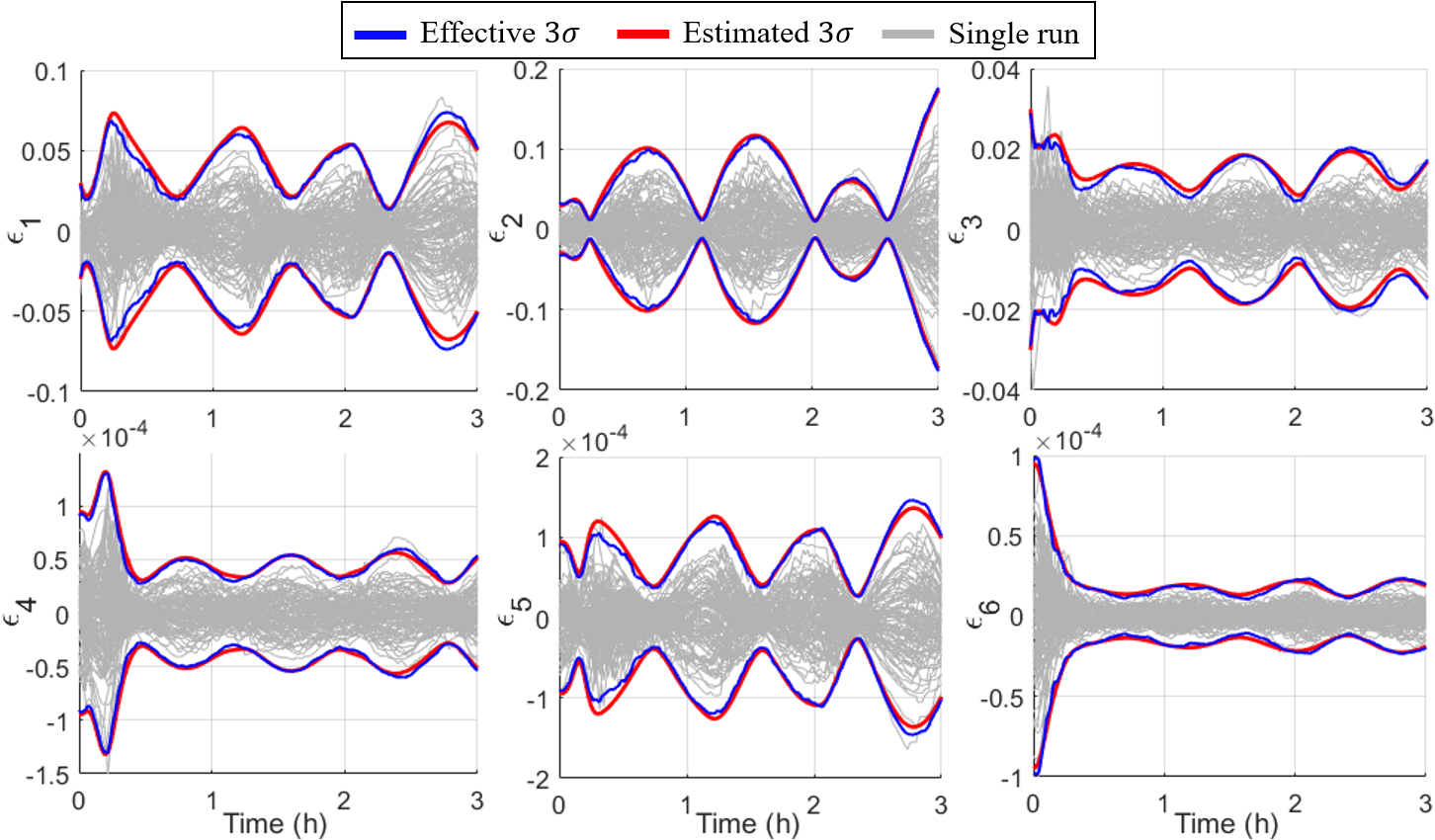}
    \caption{Monte Carlo Consistency Analysis for the Convergence of the QEKF.}
    \label{fig:mc}
\end{figure*}
The accuracy level and consistency of the filter are evaluated via a Monte Carlo analysis, reported in Fig. \ref{fig:mc} for the QEKF. In the figure, the position and velocity state errors for each run are evaluated at each time step, as
\begin{equation}
    \epsilon_i = \hat{x}_i - x_{true}
\end{equation}
Each gray line is a separate simulation. The \textit{estimated} covariance, in terms of $3\sigma$ boundary, is calculated directly from the diagonal entries of the updated covariance matrix of the filter, and reported in red. These lines indicate the filter's own estimation of its error spread and indicate how confident it is in its estimate. On the contrary, the \textit{effective} covariance level, again in terms of the $3\sigma$ boundary, represents how the filter accuracy actually performs, as these levels of standard deviation are calculated directly from the MC runs. A consistent filter has an overlapping between the estimated and the effective error covariance levels, meaning that it is able to correctly predict its own uncertainties in the estimates that it provides. Thus, the QEKF is a consistent estimator that achieves the correct tracking of the target motion. The QUKF behaves similarly to the QEKF, and, thus, the MC analysis has not been reported.

\begin{figure*}  [htbp!]
    \centering
    \includegraphics[width=.85\linewidth]{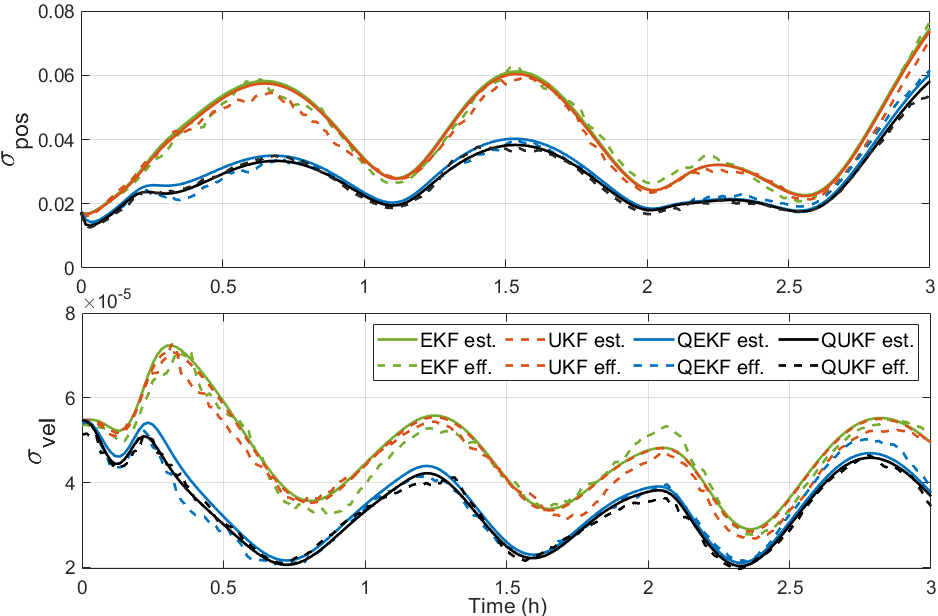}
    \caption{Error Standard Deviation Comparison between the linear estimators (EKF, UKF) and the Quadratic Estimators (QEKF, QUKF)}
    \label{fig:comp}
\end{figure*}

Now that the consistency of the QEKF and QUKF has been assessed, it is of interest to compare their accuracy to their linear counterparts. Figure \ref{fig:comp} shows the advantages in accuracy obtained by a QMMSE filter over an LMMSE one. The figure reports the error standard deviation levels both for position and velocity. Estimated covariances are evaluated directly from the updated step of the filter as 
\begin{align}
    \sigma_{pos, EST} &= \sqrt{P_{xx}+P_{yy}+P_{zz}} \\
    \sigma_{vel, EST} &= \sqrt{P_{v_xv_x}+P_{v_yv_y}+P_{v_zv_z}} 
\end{align}
while the effective error standard deviation, coming from the Monte Carlo analysis with multiple runs, is evaluated as
\begin{align}
    \sigma_{pos, EFF} &=  \sqrt{\sum_{j=\{x,y,z\}}\Bigg(    \sum_{i = 1}^{N_{MC}}(\epsilon_{j,i}- \hat{\epsilon_j})^2 \Bigg)   } \\
    \sigma_{vel, EFF} &=  \sqrt{\sum_{j=\{v_x,v_y,v_z\}}\Bigg(    \sum_{i = 1}^{N_{MC}}(\epsilon_{j,i}- \hat{\epsilon_j})^2 \Bigg)   }  
\end{align}
for each time step of the simulation. Once again, a consistent filter is assessed by the overlapping of the two standard deviations, reported as continuous and dashed lines in the figure, respectfully. At first sight, it is noted that the EKF (in green) and the UKF (in orange) have a similar accuracy level and performance. The QEKF (in blue) and the QUKF (in black) show evident benefits in accuracy, with error levels well below the other filters. The two quadratic update estimator better accounts for the non-Gaussianity of the noise distribution thanks to the introduction of the high-order central moments. Between the two QMMSE estimators, the QUKF is more accurate than the QEKF due to the use of the unscented transformation over mere linearization, as expected, and shown in the transient section of the simulation in the first hour of measurement acquisition.  

The filters proposed in this paper provide a natural improvement of the classic estimators in the sense that a parabolic fitting always provides a more accurate representation than linear regression. We were able to obtain a quadratic approximation of the MMSE using well-known uncertainty propagation techniques, such that the resulting QEKF and QUKF are easily accessible to the general reader as new filter benchmarks. Moreover, the quadratic formulation of the update, which is the core of the QMMSE, can be expanded to any uncertainty transformation technique, as the Kronecker formulations provided work regardless of how $\mbf P_{yy}$ has been evaluated.

\section{Conclusion}
The paper reported two new filtering techniques based on the QMMSE, promoting nonlinear estimators. While the QEKF and QUKF can be seen as the natural improvement of EKF and UKF, the formulation and derivations reported in the paper are expandable to any type of uncertainty quantification technique. That is, the quadratic update approximates the true MMSE more accurately than the linear update, especially when influenced by non-Gaussian noise that requires information of high-order central moments, such as skewness and kurtosis. The QUKF and QEKF obtain more information in the evaluation of the augmented Kalman gain to provide a more accurate estimate, as they know more information regarding the true shape of the posterior distribution. 

Conceptually, the technique can be expanded to include third and higher-order polynomial approximations of the update, but they require the precise and attentive evaluation of the influence of the noise on the evaluation of expected values whenever dealing with the residual. In fact, as the order of the update increases, the knowledge of higher-order central moments of distributions must be provided or approximated to guarantee the correct functionality of the filter. 

Regarding the quadratic approximations provided here, the QEKF and QUKF propose significant improvements and benefits using well-known techniques and minor changes to the original EKF and UKF algorithms, so that the benefits in accuracy come with limited added computational cost.

\bibliographystyle{ieeetr}
\bibliography{references.bib}

\end{document}